\documentclass[aps,prl,reprint,superscriptaddress,showpacs,preprintnumbers,amsmath,amssymb]{revtex4-1}
\usepackage{graphicx}  
\usepackage{dcolumn}   
\usepackage{bm}        
\usepackage{amssymb, amsmath, amsfonts}   
\usepackage{lipsum}        
\usepackage{color}
\DeclareGraphicsExtensions{{.pdf}}
\graphicspath{{./}{./figure}}
\begin{document}
\newcommand{\noter}[1]{{\color{red}{#1}}}
\newcommand{\noteb}[1]{{\color{blue}{#1}}}
\newcommand{\field}{\left( \boldsymbol{r}\right)}
\newcommand{\paren}[1]{\left({#1}\right)}
\newcommand{\vect}[1]{\boldsymbol{#1}}
\newcommand{\uvect}[1]{\tilde{\boldsymbol{#1}}}
\newcommand{\vdot}[1]{\dot{\boldsymbol{#1}}}
\newcommand{\vder}{\boldsymbol{\nabla}}
%
\widetext
%
%
\title{{Avalanche Interpretation of the Power-Law Energy Spectrum} in Three-Dimensional Dense Granular Flow}
\author{Norihiro Oyama}
\email{oyama.norihiro@aist.go.jp}
\affiliation{Mathematics for Advanced Materials-OIL, AIST, Sendai 980-8577, Japan}
\author{Hideyuki Mizuno}
\affiliation{Graduate School of Arts and Sciences, The University of Tokyo, Tokyo 153-8902, Japan}
\author{Kuniyasu Saitoh}
\affiliation{Research Alliance Center for Mathematical Sciences, Tohoku University, 2-1-1 Katahira, Aoba-ku, Sendai 980-8577, Japan}
\affiliation{WPI-Advanced Institute for Materials Research, Tohoku University, 2-1-1 Katahira, Aoba-ku, Sendai 980-8577, Japan}
\date{\today}
\begin{abstract}
  Turbulence is ubiquitous in nonequilibrium systems,
  and it has been noted that even dense granular flows exhibit characteristics that are typical of turbulent flow,
  such as the power-law energy spectrum.
  However, studies on the turbulent-like behavior of  granular flows are limited to two-dimensional (2D) flow.
  We demonstrate that the statistics in three-dimensional (3D) flow are qualitatively different from those in 2D flow.
  We also elucidate that avalanche dynamics can explain this dimensionality dependence.
  Moreover, we define clusters of collectively moving particles that are equivalent to vortex filaments.
  The clusters unveil complicated structures in 3D flows that are absent in 2D flows.
\end{abstract}
\maketitle
%
Turbulence is frequently observed in nature and represents a nonequilibrium system characterized by cascades of kinetic energy \cite{turb}.
Thus, the research on turbulent flows has been aimed at a wide range of materials,\ e.g.,\ incompressible fluids \cite{classical2,classical1,turb},
plasma \cite{plasma2,plasma1}, quantum vortices \cite{quantum1,quantum2},
bacterial suspensions \cite{bacteria_PNAS,bacteria_PRL},
and liquid crystals \cite{LC_PRL1,LC_PRL2}.
Nearly all these examples address continuum media,
and as is typical in statistical physics, the observed turbulence strongly depends on spatial dimensionality \cite{2Dturb}.

In addition to turbulent flows in continuum media,
\emph{turbulent-like} complex motions of discrete particles,
which are characteristic of disordered systems such as glass,
emulsions, colloidal suspensions, and granular materials \cite{pdf0,pdf1,pdf2,corl0,corl1,corl2,corl3,corl4,corl5,corl6,corl7,corl8},
have also been studied.
It has been widely observed that \emph{nonaffine displacements} of discrete particles exhibit vortex structures \cite{pdf0,pdf1,pdf2},
that their probability distribution functions (PDFs) show much wider tails than normal distributions \cite{pdf0,pdf1,pdf2},
and that their spatial correlations reveal collective behavior~\cite{corl0,corl1,corl2,corl3,corl4,corl5,corl6,corl7,corl8}.
In particular,
simulations \cite{spectrum,saitoh11,saitoh12}
and experiments \cite{Combe,turb_exp} have noted the similarity between the dynamics of dense granular flows and turbulence. 
However, such aspects of the {so-called \emph{granulence}\cite{spectrum}, \ i.e., \ turbulent-like statistical properties,}
have been studied extensively only in two-dimensional~(2D) systems,
and much less attention has been paid to three-dimensional~(3D) granular materials.

In this letter, we study dense granular flows in 3D space.
{In contrast to 2D flows, 3D systems exhibit localization of the spatial correlation of the nonaffine velocity field.
  Such localization leads to a positive exponent of the energy spectrum, which is inconsistent with the turbulence description.
  Instead, we propose a new physical picture from the perspective of avalanche dynamics.
  Our interpretation can explain the qualitative differences in the spatial correlation of the nonaffine velocities and the energy spectra between 2D and 3D systems.
}
Moreover, we also observe collectively moving clusters in 3D granular flow, which are unveiled by the coarse-grained vorticities.
These \emph{vortex-clusters}, which are characteristic of 3D systems and \emph{absent} in 2D systems, illustrate the complicated structures of 3D collective behavior.

\emph{Numerical methods.}---
We use molecular dynamics simulations of 3D granular particles.
To avoid crystallization, we randomly distribute a 50:50 binary mixture of $N=65536$ {(unless otherwise stated)} particles in an $L\times L\times L$ periodic box, where different kinds of particles have the same mass, $m$,
and different radii, $R_{\rm L}$ and $R_{\rm S}$ (where their ratio $R_{\rm L}/R_{\rm S}=1.4$).
The force between particles, $i$ and $j$, in contact is described by a linear spring-dashpot model~\cite{dem},\ i.e.,\ $\boldsymbol{f}_{ij}=-(k_{\rm s}\zeta_{ij}-\eta\dot{\zeta}_{ij})\uvect{r}_{ij}$ if $\zeta_{ij}>0$ and $\boldsymbol{f}_{ij}=\boldsymbol{0}$ otherwise.
Here, ${\uvect{r}}_{ij}$ is the unit vector pointing from the center of particle $j$ to particle $i$, $\zeta_{ij}\equiv R_i+R_j-r_{ij}$ represents the overlap between the particles, and $R_i$,
and $r_{ij}$ are the particle radii and the interparticle distance, respectively.
In the contact force, we introduce a spring constant, $k_{\rm s}$, and viscosity coefficient, $\eta$, such that the restitution coefficient is given by $e=\exp(-\pi/\sqrt{2mk_{\rm s}/\eta^2-1})\simeq 0.7$ \cite{dem}.
Then, we apply simple shear deformations to the system by the Lees-Edwards boundary conditions \cite{lees}.
In each time step, every particle position, $\vect{r}_i=(x_i,y_i,z_i)$ ($i=1,\dots,N$), is replaced with $(x_i+\Delta\gamma y_i,y_i,z_i)$, and equations of motion are numerically integrated with a small time increment, $\Delta t$.
Here, $\Delta\gamma$ represents a strain step such that the shear rate is defined as $\dot{\gamma}\equiv\Delta\gamma/\Delta t$.

Our system is specified by two control parameters,\ i.e.,\ the volume fraction of granular particles, $\varphi$, and the shear rate, $\dot{\gamma}$.
We estimate the jamming point as $\varphi_{\rm c}\simeq 0.64$ and fix the shear rate to $\dot{\gamma}t_0=2.5\times 10^{-5}$ with the time unit, $t_0\equiv\eta/k_{\rm s}$,
where the shear stress hardly depends on the shear rate if $\dot{\gamma}t_0\le 2.5\times 10^{-5}$ (see Supplemental Material (SM)\footnote{See Supplemental Material at [URL will be inserted by publisher] which includes Refs.~\cite{cg3,cg2,cg4}} for \emph{flow curves}).
In the following analyses, we utilize only the steady-state data, where the total strain is greater than unity.

\begin{figure}
\includegraphics[width=1\linewidth, bb=0 0 245 103]{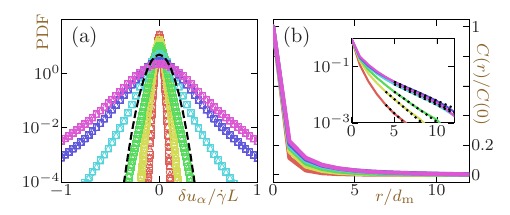}
\caption{
  {
    (a) The PDFs of nonaffine velocities, $\delta u_\alpha$ ($\alpha=x,y,z$),
where different symbols represent different components.
The dashed line is a Gaussian fit to the data for $\varphi=0.64$.
(b) Normalized spatial correlation functions, $C\paren{r}/C\paren{0}$.
The inset depicts 
semilogarithmic plots.
The dashed lines represent the exponential cutoff, $C(r)\sim e^{-r/\xi_{\rm S}}$.
Different colors indicate different values of
the volume fraction, $\varphi$, as listed in Fig.~\ref{fig:fig2}(a).
}
}
\label{fig:fig1}
\end{figure}

%
\emph{Statistical properties of nonaffine velocities.}---
To investigate the {complex motions of granular particles}, we analyze their nonaffine velocities, $\delta\vect{u}_i\equiv\vect{u}_i-\dot{\gamma}y_i\vect{e}_x$ ({defined as velocity fluctuations around the mean flow} \cite{spectrum}), where $\vect{u}_i$ and $\vect{e}_x$ are the velocity of the $i$-th particle and the unit vector parallel to the $x$-axis, respectively.
Figure \ref{fig:fig1}(a) displays the PDFs of each component of the nonaffine velocities, $\delta\vect{u}=\paren{\delta u_x,\delta u_y,\delta u_z}$, where every PDF is symmetric around $\delta u_\alpha=0$ ($\alpha=x,y,z$) and comparable, 
which means that there is no preferred direction for velocity fluctuations.
This type of isotropic behavior is common in 2D flows \cite{pdf0,pdf1,pdf2} where all particles move in plane; here, we confirm that out-of-plane motions are also equiprobable in 3D flows.
As shown in Fig.\ \ref{fig:fig1}(a), the PDFs become broader with increasing $\varphi$ and significantly deviate from the normal distribution (the dashed line).

The wide tails of the PDFs suggest strong correlations between the nonaffine velocities.
To quantify such correlations between nonaffine velocities,
we introduce spatial correlation functions as $C\paren{\vect{r}}=\langle\delta\vect{u}\paren{\vect{r}}\cdot\delta\vect{u}\paren{\vect{0}}\rangle/\langle\delta\vect{u}\paren{\vect{0}}\cdot\delta\vect{u}\paren{\vect{0}}\rangle$ \cite{corl2,corl3},
where the nonaffine velocity field is defined as $\delta\vect{u}\paren{\vect{r}}\equiv\sum_i\delta\vect{u}_i\delta(\vect{r}-\vect{r}_i)$ with the Dirac delta function, $\delta(\vect{r})$.
The ensemble average, $\langle\cdots\rangle$, is taken over time and different positions for the origin, $\vect{0}$.
Because the nonaffine velocities are confirmed to be isotropic (Fig.\ \ref{fig:fig1}(a)), we further take spherical averages such that the correlation functions are given by functions of the distance,\ i.e.,\ $C\paren{r}$ with $r\equiv |\vect{r}|$.
{Figure \ref{fig:fig1}(b) displays normalized correlation functions, where the distance is scaled by the mean particle diameter, $d_{\rm m}\equiv R_{\rm L}+R_{\rm S}$.
Contrary to the expectation from the wide tails of the PDF, the correlation functions are localized.
This tendency is qualitatively different from that in the 2D system, where the spatial correlation of  nonaffine velocities is long range.
We will discuss the cause of this localization later.
The inset of Fig.~\ref{fig:fig1}(b) depicts semilogarithmic plots of the same data.
Here, we describe their tails by an exponential cutoff, $C(r)\sim e^{-r/\xi_{\rm S}}$ (the dotted lines), to introduce a correlation length as $\xi_{\rm S}$.
We show the dependence of $\xi_{\rm S}$ on the volume fraction, $\varphi$, in the inset of Fig.\ \ref{fig:fig4}(a). If the system is below jamming, $\varphi<\varphi_{\rm c}$,
the correlation length increases with the volume fraction, while it saturates in the yielding states, $\varphi>\varphi_{\rm c}$, as observed in 2D flows~\cite{pdf1,corl2,corl3,saitoh11,saitoh12}.
 We note that the saturation of the correlation length has been theoretically predicted~\cite{Berzi2015a,Berzi2015b}.
}

\begin{figure}
\includegraphics[width=1\linewidth, bb=0 0 245 196]{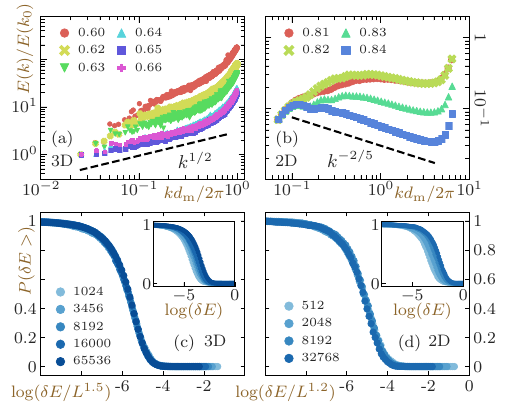}
\caption{
  {
(a) The energy spectra, $E\paren{k}/E\paren{k_0}$, where $k_0$ denotes the smallest wave number, $k_0=2\pi/L$.
The dashed line depicts the power law $E\paren{k}\sim k^{1/2}$.
(b) The cumulative probability distribution of
the magnitude (logarithmic) of the potential energy drop events.
Different colors indicate different system sizes, $N$.
The main panel shows the data collapsed by finite size scaling, while the inset shows the raw data.
(c)(d) The same plots as (a) and (b) for 2D systems. Note that the exponents of the power law range of the energy spectrum and the finite size scaling are different from those in 3D systems.
  }
}
\label{fig:fig2}
\end{figure}

\emph{Energy cascades in 3D granular flows.}---
{The Fourier space counterpart of the velocity spatial correlation function} $C\paren{\vect{r}}$, 
\ i.e.,\ the spectrum of nonaffine velocities, $e\paren{\vect{k}}= (\rho_0/2)\left\langle|\delta\hat{\vect{u}}\paren{\vect{k}}|^2\right\rangle$,
   {enables us to more directly examine energy cascades.}
Here, $\rho_0\equiv mN/L^3$ is the mass density,
and the Fourier transform of nonaffine velocities is given by $\delta\hat{\vect{u}}\paren{\vect{k}}=\sum_i\delta{\vect{u}}_i\text{e}^{-I\vect{k}\cdot\vect{r}_i}$ with the wavenumber vector, $\vect{k}$,
and imaginary unit, $I$.
The spectrum, $e\paren{\vect{k}}$, is an analog of the energy spectrum \cite{spectrum}, and its integral over the whole $\vect{k}$-space gives the \emph{granular temperature} \cite{kinetic0}.
Because the nonaffine velocities are isotropic, we also take a spherical average of the spectrum {such that $E(k)\equiv 4\pi k^2e(|\vect{k}|)$, with $k=|\vect{k}|$.
Here, we introduced the Jacobian $4\pi k^2$ to make $E(k)$ equivalent to the function used in the field of fluid turbulence.}
Figure \ref{fig:fig2}(a) shows our results for the spectra.
As seen in the figure, the spectrum exhibits power-law behavior,
  $E\paren{k}\sim k^{1/2}$, if the system is in the yielding states,
  $\varphi>\varphi_{\rm c}$, while it grows more strongly below jamming, $\varphi<\varphi_{\rm c}$
  {(we confirmed that all power-law behaviors reported in this letter do not depend on the system size; see SM \cite{Note1}).}
We find that the range of the power law is more than one decade from the system size, $k= 2\pi/L$, to about three particle diameters, $2\pi/3d_{\rm m}$.
{The energy spectrum of the conventional fluid turbulence shows power-law decay as $E(k)\sim k^{-\alpha}$ with a positive $\alpha$.
   In 2D dense granular flow, similar power-law decay of energy spectra has been reported, and the statistical analogy to conventional turbulence has been discussed  \cite{spectrum,saitoh12}. 
   On the other hand, in 3D systems, the energy spectra show a positive exponent (Fig.~\ref{fig:fig2}(a)), and the statistics of
the dynamics are qualitatively different from those of conventional turbulence.
   In other words, the \emph{granulence} is missing in 3D systems.
}

{
  \emph{Avalanche interpretation of energy spectra.}---
Then, how can the power-law behavior of the energy spectrum in granular flows be interpreted?
    We demonstrate that the interpretation from the perspective of avalanche dynamics 
    can explain the exponent in both 2D and 3D systems.
    In refs.~\cite{Maloney2004,Bailey2007}, the authors investigated  
    intermittent drops in the time evolution of the potential energy
    in amorphous solids under quasi-static steady shear,
    which are one typical feature of avalanche dynamics.
    The authors have reported that a plastic event associated
    with a single energy drop
    forms a fractal structure in both 2D and 3D systems.
    The authors clarified that the fractal dimension of such a collective rearrangement
    can be quantified by 
    the finite-size scaling exponent of the PDF of the potential energy drop sizes.
    We conducted the same finite-size scaling analysis (FSSA) in our system.
    In Fig.~\ref{fig:fig2}(b), the cumulative probability distribution of
    the energy drop is shown for various system sizes as listed in the legend.
    In the main panel, the energy drop values are scaled by $L^{1.5}$,
    and the inset shows the unscaled results.
    
    These energy drops can be related to the nonaffine velocity field.
    A part of the potential energy stored by the affine deformation 
    is released as the nonaffine velocity field when plastic rearrangements occur.
    Therefore, the spatial correlation of the nonaffine velocity fields should reflect
    the fractal structure of the plastic rearrangements.
    Based on this physical picture, we can write the energy balance equation among
    the energy released due to the rearrangements
    ${\epsilon}(r)$, the kinetic energy field of the nonaffine velocity
    $e(r)$ and the local energy dissipation field
    $\chi(r)$ as ${\epsilon}(r)\simeq e(r) + \chi (r)$.
    The FSSA states that the integral of the energy release distribution gives the following relation, $\int_0^r\epsilon(r^{\prime})r^{\prime2} dr^\prime \propto r^{3/2}$,
    which can be reduced to $\epsilon(r)\propto r^{-3/2}$.
    Because the dissipation is negligible at a large wavelength scale,
    we obtain $\hat{\epsilon}(k)\sim\hat{e}(k)\propto k^{-3/2}$, where $\hat{\epsilon}(k)$ and $\hat{e}(k)$ are the Fourier transforms of $\epsilon(r)$ and $e(r)$, respectively.
    Finally, this gives the scaling relation $E(k)=4\pi k^2 \hat{e}(k)\propto k^{1/2}$
    for small values of $k$.

    {To show that this avalanche interpretation can provide a unified physical picture for power-law energy spectra in both 2D and 3D systems, we have also conducted the same analyses in 2D systems.
      We performed simulations of 2D dense granular flow
      with the same parameters as those used for 3D systems (only the densities are different due to the dimensionality difference).
      As a result, we obtain the relation $E(k)\sim k^{-1/5}$ from the FSSA and $E(k)\sim k^{-2/5}$ from the direct measurement of the energy spectrum, as shown in Figs.~\ref{fig:fig2}(c,d).
      These values of exponents are reasonably close; thus, we conclude that the power-law exponent of the energy spectrum can be understood as a reflection of the underlying avalanche dynamics.}
    Similarly, because the spatial correlation of nonaffine velocities
    is governed by the fractal structure of plastic events,
    the qualitative difference in the energy spectrum can provide an intuitive explanation
    of the localization of the velocity correlation in 3D systems, while it becomes long-ranged in 2D flow.
We stress that this avalanche interpretation can also explain why we observe these nontrivial statistics only under a very low shear rate and above jamming.
This regime is exactly where we observe jerky time evolutions of macroscopic values reflecting avalanches.
}

\begin{figure}
\includegraphics[width=\linewidth, bb=0 0 1942 799]{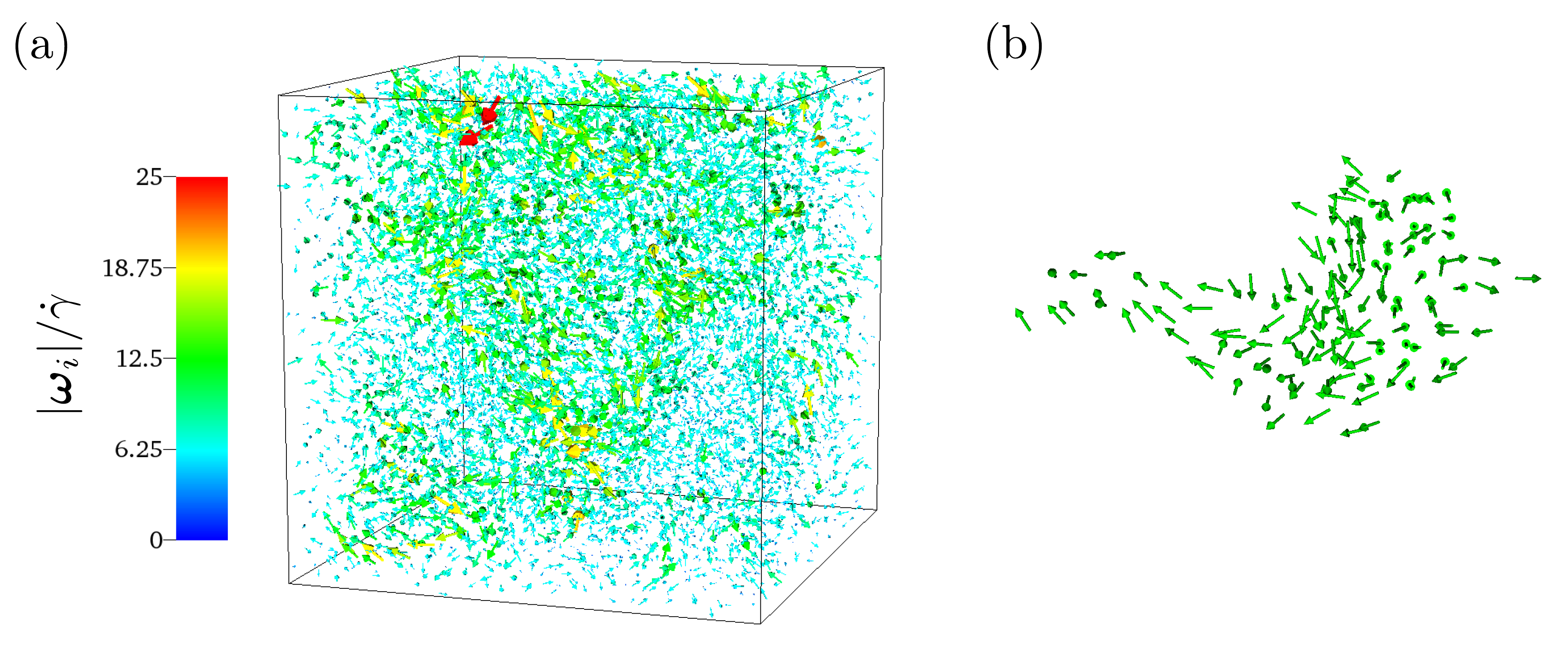}
\caption{\label{fig:vis}
{(a) The vorticity at each particle position (arrows),
  where the color represents the normalized magnitude.
  (b) A multiaxial vortex-cluster of size $N_{\rm C}=141$.
  All vectors are illustrated in the same color.
In both (a) and (b), the volume fraction and shear rate are given by $\varphi=0.66$ and $\dot{\gamma}t_0=2.5\times10^{-5}$, respectively.}
}
\end{figure}

\begin{figure}
\includegraphics[width=0.8\linewidth, bb=0 0 245 306]{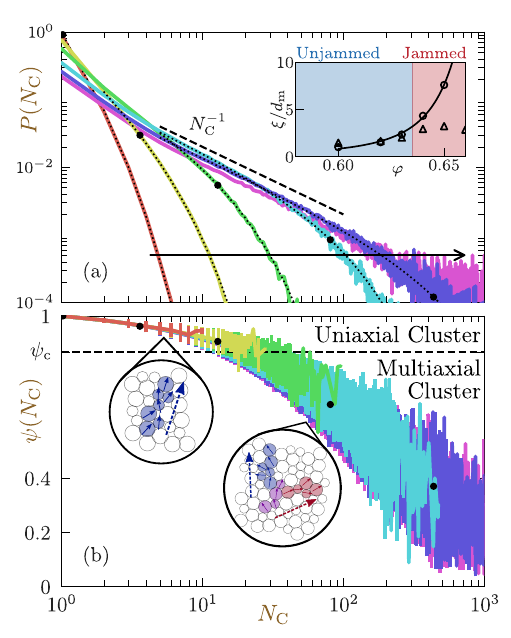}
\caption{
  {
(a) The PDFs of the cluster size $N_{\rm C}$.
The dashed line depicts the power-law decay $P\paren{N_{\rm C}}\sim N_{\rm C}^{-1}$,
and the dotted lines represent the fitting curves.
The volume fraction $\varphi$ increases,
as indicated by the arrow.
The inset shows two distinct characteristic lengths, $\xi_{\rm C}$ (circles) and $\xi_{\rm S}$ (triangles), as functions of the volume fraction.
The solid line in the inset is introduced as a guide for the eye.
(b) The vorticity polar order parameter $\psi\paren{N_{\rm C}}$
of clusters as a function of $N_{\rm C}$.
The dashed line indicates the critical value $\psi_{\rm c}$.
The inset sketches are schematic images of uniaxial and multiaxial clusters.
Different colors indicate different values of
the volume fraction, as listed in Fig.~\ref{fig:fig2}(a).
The filled circles indicate the characteristic cluster sizes $\nu_{\rm C}$.}
  }
\label{fig:fig4}
\end{figure}

\emph{Analyses of vortex-clusters.}---
{One major difference between 2D and 3D systems is the degree of freedom of vorticities.
  While vorticities are all parallel to the out-of-plane $z$-axis~\cite{saitoh12} in the 2D case,
  vorticities can rotate freely in 3D systems, and their 3D structures,\ e.g.,\ \emph{vortex filaments}, are a unique feature of this dimension.}
Therefore, we next turn our attention to the vorticity field,\ i.e.,\ $\vect{\omega}(\vect{r})=\vect{\nabla}\times\delta\vect{u}(\vect{r})$, {which will unveil the complicated and long-ranged structure of the collectivity in 3D granular flows. }
Here, we calculate the vorticity field by the coarse-graining (CG) method as \cite{cg2,cg4}
\begin{equation}
\vect{\omega}\field = \rho\field^{-2}\sum_{i=1}^{n}\sum_{j=1}^n\Psi_j\field\left[ \vder\Psi_i\field\times\delta\vect{u}_{ij}\right]~,
\label{eq:vorticity}
\end{equation}
where $\rho\field$ and $\delta\vect{u}_{ij}=\delta\vect{u}_i-\delta\vect{u}_j$ are the number density field and the relative nonaffine velocity between the particles, $i$ and $j$, respectively.
On the right-hand-side of Eq.\ (\ref{eq:vorticity}), the CG kernel for the $i$-th particle is given by a Gaussian function, $\Psi_i\field\equiv e^{-|\vect{r}-\vect{r}_i|^2/d_{\rm m}^2}$ (see SM \cite{Note1} for full details).
{Figure \ref{fig:vis}(a) displays a snapshot of the vorticity at each particle position,\ i.e.,\ $\vect{\omega}_i\equiv\vect{\omega}\paren{\vect{r}_i}$,
  where the volume fraction is given by $\varphi=0.66>\varphi_{\rm c}$.
In this figure, we observe that the vorticity distributes heterogeneously in space and that there is no preferred direction, as in the case of nonaffine velocities (Fig.\ \ref{fig:fig1}(a)).}
In the SM \cite{Note1}, we show that the PDFs of vorticity are symmetric around zero and exhibit wide tails.
To detect 3D structures of vorticity, we identify clusters by a similar method to that used for Janus particles \cite{janus1,janus2}: If two particles, $i$ and $j$,
are in contact, we compute the angle between $\vect{\omega}_i$ and $\vect{\omega}_j$ as $\theta_{ij}\equiv\cos^{-1}\paren{\uvect{\omega}_i\cdot\uvect{\omega}_j}$,
where $\uvect{\omega}_i\equiv\vect{\omega}_i/|\vect{\omega}_i|$ represents the direction of vorticity.
Then, if the angle is smaller than a threshold,\ i.e.,\ $\theta_{ij}<\theta_{\rm c}$, we include the particles in the same cluster.
We call these clusters \emph{vortex-clusters}.
Figure \ref{fig:vis}(b) shows a snapshot of a vortex-cluster, where we choose $\theta_{\rm c}=\pi/6$ as the threshold
\footnote{The choice of $\theta_{\rm c}$ is arbitrary, and the cluster analysis is sensitive to $\theta_{\rm c}$. For example, if $\theta_{\rm c}$ is too small,
  no cluster is detected, and if $\theta_{\rm c}$ is too large, every particle belongs to a single cluster.
  We chose $\theta_{\rm c}=\pi/6$ so that $\xi_{\rm C}$ converges to $\xi_{\rm S}$ below jamming.}.
As can be seen, the 3D structures of vorticity are nicely visualized, and their bending, twisting, and branching shapes, which cannot be detected by the spatial correlation functions (Fig.\ \ref{fig:fig1}(b)), are well captured by vortex-clusters.

We quantify the size of vortex-clusters to explain collective motions above jamming.
Figure \ref{fig:fig4}(a) displays the PDFs of cluster size, $P\paren{N_{\rm C}}$, where $N_{\rm C}$ is the number of particles in a vortex-cluster.
In this figure, all the tails are well described by the power law with an exponential cutoff,\ i.e.,\ $P\paren{N_{\rm C}}\sim N_{\rm C}^{-\alpha}{\rm exp}\paren{-N_{\rm C}/\nu_{\rm C}}$.
Similar PDFs of cluster size have been reported experimentally for 2D granular system where clusters are defined in terms of the dynamic heterogeneity \cite{exp1}.
Therefore, by adjusting the exponent, $\alpha$, and characteristic size, $\nu_{\rm C}$, we estimate a characteristic length scale from $\xi_{\rm C}\equiv \nu_{\rm C}^{1/3}$
(indicated by the filled circles in Fig.~\ref{fig:fig4}(a)).
We find that the exponent slightly decreases with increasing volume fraction below jamming, $\varphi<\varphi_{\rm c}$,
while it converges to $\alpha\simeq 1$ above jamming, $\varphi>\varphi_{\rm c}$.
On the other hand,
the characteristic size, $\nu_{\rm C}$, increases above jamming such that the length scale, $\xi_{\rm C}$, does not plateau in $\varphi>\varphi_{\rm c}$ (the inset of Fig.\ \ref{fig:fig4}(a)).
This is a remarkable difference from the correlation length $\xi_{\rm S}$ defined by $C(r)\sim e^{-r/\xi_{\rm S}}$.
We note that $\nu_{\rm C}$ reaches the system size, $N$, at $\varphi=0.66$, \ i.e.,\ the vortex-clusters extend over the whole system.

\emph{Structure of vortex-clusters.}---
The structural change in the yielding state
can be revealed by the vorticity polar order parameter $\psi_k=\frac{1}{N_k}\left| \sum_i  \uvect{\omega}_i\right|$ \cite{PO1,PO2}, where $N_k$ is the number of particles in the $k$-th cluster.
When all vorticities are completely parallel, we obtain $\psi_k=1$, and when the directions are randomly distributed, we expect to have $\psi_k\simeq 1/\sqrt{N_k}$ \cite{PO2}.
In addition to these common features, in the current situation, a critical value $\psi_{\rm c}=\text{cos}\theta_{\rm c}$ distinguishes qualitatively different structures.
If  $\text{cos}^{-1}\paren{\uvect{\Omega}_k\cdot \uvect{\omega}_i} < \theta_{\rm c} (i=1,2,\cdots N_k)$ is satisfied for all $i$,
then $\psi_k$ is larger than this critical value $\psi_{\rm c}$,
where $\uvect{\Omega}_k$ is the unit vector parallel to the mean vorticity $\vect{\Omega}_k\equiv \frac{1}{N_k}\sum_i^{N_k}\uvect{\omega}_i$ (see SM \cite{Note1} for full details of $\psi_{\rm c}$).
Because this cluster is characterized by a single axis, $\uvect{\Omega}_k$, we name it a \emph{uniaxial cluster}.
On the other hand,  $\psi_k$ is expected to be less than $\psi_{\rm c}$ for the case of \emph{multiaxial clusters}, where multiple uniaxial clusters are connected by bridging particles between individual uniaxial clusters (purple particles in Fig.~\ref{fig:fig4}~(b) inset).
We stress that such multiaxial clusters cannot be realized in 2D systems, where all vorticity vectors must be parallel.
The size dependence of the mean polar order $\psi\paren{N_{\rm C}}\equiv \langle \psi_k \delta_{{N_{\rm C}, N_k}}\rangle$ is shown in Fig.~\ref{fig:fig4}~(b), where $\delta_{\alpha,\beta}$ is the Kronecker delta and the average $\langle \cdots \rangle$ is taken over all clusters.
As a result, we notice that the characteristic cluster sizes $\nu_{\rm C}$ for the unjammed regime ($\varphi<\varphi_{\rm c}$) all meet $\psi\paren{\nu_{\rm C}}>\psi_{\rm c}$.
In other words, most clusters in the unjammed regime are uniaxial.
On the other hand, in the jammed regime, $\psi\paren{\nu_{\rm C}}$ becomes less than $\psi_{\rm c}$, which means that multiaxial clusters are observed rather frequently.
Note that different branches of multiaxial clusters are considered uncorrelated by the spatial correlation function $C\paren{r}$, while within a branch or a uniaxial cluster, $C\paren{r}$ can evaluate the correlation properly.
 {We consider this change in morphology at the jamming transition to lead to pronounced discrepancies between $\xi_{\rm S}$ and $\xi_{\rm C}$ only in the jammed regime.
   }

\emph{Summary.}---
To conclude, {we confirmed that so-called \emph{granulence} is absent in 3D systems.}
{The absence is evidenced by the localization of the correlation of the nonaffine velocity field
  and more directly by the positive exponent of the energy spectrum.
  Although, from the perspective of  \emph{granulence},
  2D and 3D systems seem qualitatively different,
  we propose an interpretation from the perspective of avalanche dynamics that provides
  a unified  understanding of the results in different dimensions.
}
      {Furthermore, by defining vortex-clusters with respect to the coarse-grained vorticities, we illustrated collectivity in 3D granular flows.
We emphasize that these vortex-clusters are characteristic of 3D systems and can\textit{not} exist in 2D flows.
We also found that the vortex-clusters show complicated, multiaxial structures in the yielding states.}

\begin{acknowledgments}
We thank {Michio Otsuki, John J.\ Molina, Simon K.\ Schnyder, Hiroaki Itoh, Satoru Tokuda and  Masanari Shimada} for useful discussions.
This work was financially supported by the WPI, MEXT, Japan,
a Grant-in-Aid for Scientific Research B (No. 16H04025), a Grant-in-Aid for Young Scientists B (No. 17K14369),
and a Grant-in-Aid for Early-Career Scientists (No. 18K13464) from the Japan Society for the Promotion of Science (JSPS).
\end{acknowledgments}

%
%


%
%
\end{document}


\renewcommand{\theequation}{SM \arabic{equation}}
\renewcommand{\thesection}{SM \arabic{section}}
\renewcommand{\thefigure}{SM\arabic{figure}}

\section{The rheological property of the system}
The macroscopic rheological property of dense granular flow is specified by two control parameters, $i.e.$, the steady shear rate $\dot{\gamma}t_{\rm m}$
and the volume fraction $\varphi$.
In this Supplemental Material,
we present the dependence of the rheological property on these two control parameters.

First, we define the macroscopic stress tensor as
\begin{align}
  \sigma_{\alpha\beta}=\frac{1}{L^3}\left< \left( \sum_i \sum_{j\neq i}f_{ij \alpha}r_{ij \beta}
  + \sum_i m \delta u_{i\alpha} \delta u_{i\beta}\right)\right> \label{eq:stress}\\
  \left(   \alpha, \beta \in \left\{x, y, z\right\} \text{ and } i,j \in \left\{ 1,2,\cdots N\right\}\right)\notag
\end{align}
where $\delta u_{i\alpha}$, $f_{ij\alpha}$, and $r_{ij\alpha}$ represent the $\alpha$ component of the nonaffine velocity $\delta\boldsymbol{u}_i$,
the interparticle contact force $\boldsymbol{f}_{ij}$, and the relative position $\boldsymbol{r}_{ij}$, respectively.
The first term of the right-hand side of Eq.~(\ref{eq:stress}) reflects the kinetic contribution to the stress, and the second term reflects the contact contribution.
As a measure of the rheology, we use
$\sigma=\langle \frac{1}{2}\left( \sigma_{xy}+\sigma_{yx}\right)\rangle,$
which is the direct mechanical response to the external shear.
Note that the angular bracket here means the average over the values in the steady state $\gamma > 1$.

\begin{figure}[hb]
  \begin{center}
    \includegraphics[width=\linewidth, bb=0 0 510 192]{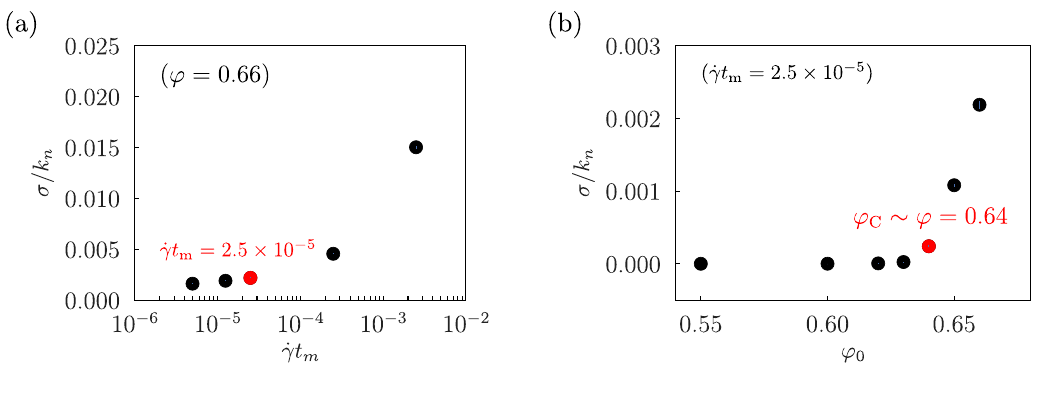}
    \caption{\label{fig:ss_sr}
      The mean shear stress in the units of normal spring constant, $\sigma/k_{\rm n}$, as functions of control parameters, (a) $\dot{\gamma}t_{\rm m}$ and (b) $\varphi$.
      (a) All data are for the case with ${\varphi}=0.66$. The value used in the calculations in the main text is marked in red. (b) All data are for the case with $\dot{\gamma}t_{\rm m}=2.5\times 10^{-5}$. The jamming transition point is marked in red.
}
\end{center}
\end{figure}

In Fig.~\ref{fig:ss_sr}~(a), the shear rate $\dot{\gamma}t_{\rm m}$ dependence of the mean shear stress $\sigma$ is shown for a  single given value of volume fraction $\varphi=0.66$.
The shear stress $\sigma$ hardly depends on the shear rate below $\dot{\gamma}t_{\rm m}=2.5\times 10^{-5}$, which we use for all the results in the main text.
In Fig.~\ref{fig:ss_sr}~(b), the volume fraction $\varphi$ dependence of $\sigma$ is shown for a  single given value of shear rate $\dot{\gamma}t_{\rm m}=2.5\times 10^{-5}$.
A transition from the unjammed regime to the jammed regime can be observed around the critical volume fraction $\varphi_{\rm C}\sim 0.64$, which is indicated by a sharp growth in the shear stress.

\newpage
\section{Localization of the spatial correlation of the nonaffine velocity field in 3D}\label{sec:localize}
The spherically averaged spatial correlation function decays very fast in 3D dense granular flow,
while it shows a long-range correlation in 2D.
The avalanche fractal dimension of the energy release distribution $\epsilon(r)$ due to plastic events (introduced in the main text) can also provide a possible scenario for such a localization of the correlation in 3D systems.

We can estimate the volume of the activated region in a shell of width $dr$ separated by $r$ from an arbitrary activated point as $C_E(r)\propto\epsilon(r)J(r)dr$, where $J(r)$ is the Jacobian.
If we take the spherical average, we obtain $C_E^{\rm iso}(r)\propto \epsilon(r)dr$.
This function $C_E^{\rm iso}$ again shows a dimensionality-dependent qualitative difference as:
\begin{align}
  C_{E,2D}^{\rm iso}&\propto r^{1/5},\\
  C_{E,3D}^{\rm iso}&\propto r^{-1/2}.
\end{align}
This function $C_E^{\rm iso}$ detects only the energetic correlation and
does not count the vectorial information of nonaffine velocities.
In this sense, this function can be understood as the upper bound of the velocity correlation.
The negative power results in the localized correlation
even with respect to the upper bound.
Note that if we calculate the directional correlation instead of taking the spherical average, we do observe long-range correlation even in 3D.
That is what we observed by vortex cluster analysis.

\newpage
\section{Check for the finite-size effect on the power-law behaviors}
\begin{figure}[htbp]
  \begin{center}
    \begin{tabular}{c}

      \begin{minipage}{0.5\hsize}
        \begin{center}
          \includegraphics[clip, width=0.8\linewidth, bb=0 0 245 306]{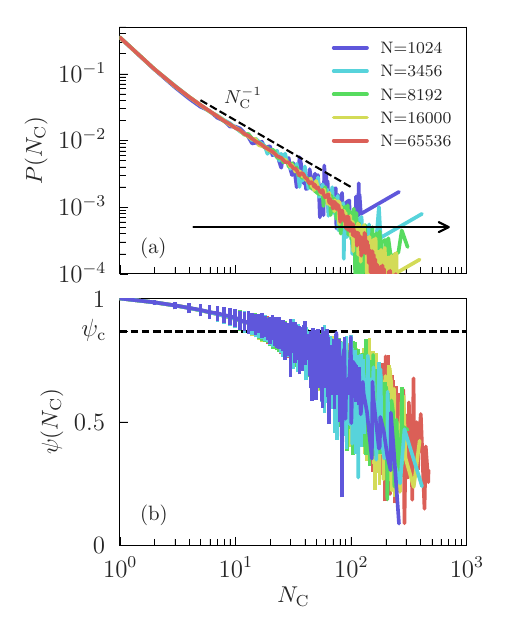}
          \hspace{1.6cm} 
        \end{center}
      \end{minipage}

      \begin{minipage}{0.5\hsize}
        \begin{center}
          \includegraphics[clip, width=0.8\linewidth, bb=0 0 245 184]{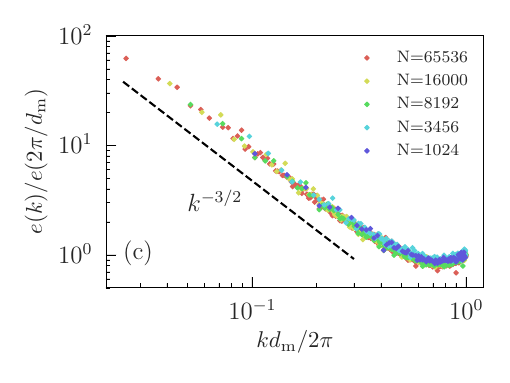}
          \hspace{1.6cm} 
        \end{center}
      \end{minipage}

    \end{tabular}
    \caption{The system size dependence of (a) the cluster size distribution, (b) the polar order parameter and (c) the energy spectrum density $e(k)$.
 The values in (c) are normalized by the value at $k=2\pi/d_{\rm m}$, which corresponds to the mean particle diameter.
All figures show very good overlaps without any factor or scaling.}
    \label{fig:FSS}
  \end{center}
\end{figure}

We have checked the finite-size effect on the functions that exhibit power-law behaviors, namely, the energy spectrum (density) and the cluster size distibution.
To this end, we have conducted simulations of systems with $N=1024, 3456, 8192, 16000$ and $65536$ (the largest value is the one used in the main text).
As shown in Fig.~\ref{fig:FSS}, 
the results show beautiful superposition without any factors, and we could confirm that none of the power-law behaviors observed in this work depend on the system size.

\newpage
\section{Coarse-grained vorticity field}
We briefly explain the definition of the \emph{coarse-grained vorticities}.
To define continuum field variables in discrete systems,
we employ an unscaled Gaussian function as the coarse-graining kernel~\cite{cg3,cg2}
\begin{align}
  \Psi \left( \boldsymbol{r}\right)=e^{-\left( \boldsymbol{r}/d_{\rm m}\right)^2},
\end{align}
where $\boldsymbol{r}$ is an arbitrary position in the computational domain and $d_{\rm m}$ is the mean particle diameter.
In this coarse-graining description, the number density field and nonaffine velocity field can be formulated as~\cite{cg3,cg2}
\begin{align}
  \rho \left( \boldsymbol{r}\right) &= \sum_{i=1}^N \Psi \left( \boldsymbol{r}-\boldsymbol{r}_i\right),\\
  \delta \boldsymbol{u} \left( \boldsymbol{r}\right)
  &= \rho \left( \boldsymbol{r}\right)^{-1}\sum_{i=1}^N \delta \boldsymbol{u}_i\Psi \left( \boldsymbol{r}-\boldsymbol{r}_i\right).
\end{align}
Using this interpolated nonaffine velocity field, we can then define the derivative of the $\alpha$ component of the nonaffine velocity field $\delta u_\alpha\left(\boldsymbol{r} \right)$ with respect to the $\beta$ component of position $r_{\beta}$ as
\begin{align}
  \frac{\partial \delta u_{\alpha}}{\partial r_{\beta}}=\rho \left( \boldsymbol{r}\right)^{-2}
  \sum_{i=1}^{N}\sum_{j=1}^{N}\delta u_{ij\alpha} \Psi_j
  \frac{\partial\Psi_i}{\partial r_{\beta}},\label{eq:der}
\end{align}
where $\delta_{ij\alpha}$ denotes the $\alpha$ component of the relative nonaffine velocities $\delta\boldsymbol{u}_{ij}\equiv \delta\boldsymbol{u}_i-\delta\boldsymbol{u}_j$
and $\Psi_i$ is a abbreviation for the coarse-graining kernel due to particle $i$, $\Psi\left( \boldsymbol{r}-\boldsymbol{r}_i\right)$.
Using Eq.~(\ref{eq:der}), we finally derive the coarse-grained vorticity field as~\cite{cg4}
\begin{align}
  \boldsymbol{\omega} \left( \boldsymbol{r} \right)
  =\rho\left( \boldsymbol{r}\right)^{-2}\sum_{i=1}^{N}\sum_{j=1}^N\Psi_j\left( \boldsymbol{\nabla}\Psi_i \times \delta \boldsymbol{u}_{ij}\right).
\end{align}
The coarse-grained vorticity of particle $i$ is defined as $\boldsymbol{\omega}_i=\boldsymbol{\omega}\left( \boldsymbol{r}_i\right)$.
\newpage
\section{The probability distribution of vorticities}
\begin{figure}[hb]
\includegraphics[width=0.5\linewidth, bb=0 0 245 184]{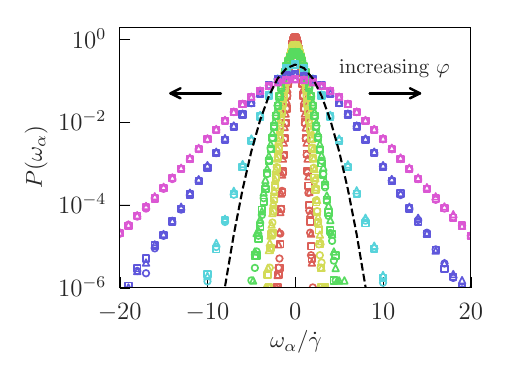}
\caption{(Color online)
Semilogarithmic plots of the PDFs of vorticities, $P(\omega_\alpha)$ ($\alpha=x,y,z$),
where the nonaffine vorticities are scaled by the shear rate as $\omega_\alpha/\dot{\gamma}$
and different symbols,\ i.e.,\ squares, circles, and triangles, represent different components, $\omega_x$, $\omega_y$, and $\omega_z$, respectively.
The volume fraction increases from $\varphi=0.60$ to $0.66$ (as indicated by the arrows), and the color code is the same as in Fig.\ 2 in the main text.
The dashed line is a Gaussian fit to the data for $\varphi=0.64$.
\label{fig:PDF}}
\end{figure}

The probability distribution function (PDF) of each component of the vorticities, $\omega = \left( \omega_x, \omega_y, \omega_z \right)$, is shown in Fig.~\ref{fig:PDF}.
As is the case for PDFs of the nonaffine velocities, every PDF is symmetric around $\omega_\alpha = 0 \left( \alpha=x,y,z\right)$ and overlaps completely with other PDFs from the same volume fraction $\varphi$.
The volume fraction $\varphi$ dependence is also consistent with that of nonaffine velocities: PDFs become broader with the increase in $\varphi$, and the non-Gaussianity is remarkable for the cases with high values of $\varphi$.

\newpage
\section{The meaning of the critical polar order $\psi_{\rm c}$}
Here, we show how the threshold angle $\theta_{\rm c}$ for the vortex-cluster definition can lead to the critical polar order $\psi_{\rm c}$, which distinguishes uniaxial and multiaxial clusters.

As introduced in the main text, the polar order is given as
\begin{align}
  \psi_k\equiv\left|  \frac{1}{N_k}\sum_i^{N_k}\tilde{\boldsymbol{\omega}}_i \right|=\left| \boldsymbol{\Omega}_k\right|,\label{eq:polar}
\end{align}
where  ${\boldsymbol{\Omega}}_k=\frac{1}{N_k}\sum_i^{N_k}\tilde{\boldsymbol{\omega}_i}$.
Therefore, the unit vector parallel to $\boldsymbol{\Omega}_k$ can be expressed as $\tilde{\boldsymbol{\Omega}}_k=\boldsymbol{\Omega}_k/\psi_k$.

Squaring the both sides of Eq.~\ref{eq:polar}, we can remove the absolute value and obtain
\begin{align}
  \psi_k^2 &= \frac{1}{N_k^2}\left| \sum_k^{N_k}\boldsymbol{\omega}_i\right|^2\notag\\
  &=\frac{1}{N_k^2}\sum_i^{N_k}\sum_j^{N_k}\left( \boldsymbol{\omega}_i\cdot \boldsymbol{\omega}_j\right)\label{eq:sq}
\end{align}

On the other hand, the inner product of $\tilde{\boldsymbol{\Omega}}_k$ and $\tilde{\boldsymbol{\omega}}_i$ can be reduced as
\begin{align}
  \tilde{\boldsymbol{\Omega}}_k\cdot \tilde{\boldsymbol{\omega}}_i
  &= \frac{\boldsymbol{\Omega}_k}{\psi_k}\cdot \tilde{\boldsymbol{\omega}}_i\notag\\
  &= \frac{1}{\psi_k N_k}\sum_j^{N_k}\tilde{\boldsymbol{\omega}}_j\cdot\tilde{\boldsymbol{\omega}}_i\notag\\
  \therefore \sum_i^{N_k}\tilde{\boldsymbol{\Omega}}_k\cdot\tilde{\boldsymbol{\omega}}_i
  &= \frac{1}{\psi_k N_k}\sum_i^{N_k}\sum_j^{N_k}\left(
  \tilde{\boldsymbol{\omega}}_i\cdot\tilde{\boldsymbol{\omega}}_j\right)\label{eq:inpr}
\end{align}
Substituting Eq.~\ref{eq:inpr} into Eq.~\ref{eq:sq}, we obtain
\begin{align}
  \psi_k^2 &= \frac{1}{N_k}\psi_k\sum_i^{N_k}\tilde{\boldsymbol{\Omega}}_k\cdot\tilde{\boldsymbol{\omega}}_i\notag\\
  &=\frac{1}{N_k}\psi_k\sum_i^{N_k}\text{cos}\theta_i\notag\\
  \therefore \psi_k &= \frac{1}{N_k}\sum_i^{N_k}\text{cos}\theta_i,
\end{align}
where we introduced $\theta_i$ such that $\tilde{\boldsymbol{\Omega}}_k\cdot\tilde{\boldsymbol{\omega}}_i=\text{cos}\theta_i$.
Thus, the polar order parameter $\psi_k$ can be described in terms of the average of $\text{cos}\theta_i$.
Therefore, if $\theta_i<\theta_{\rm c}$ holds for all $i\left( i=1,2,\cdots N_k\right)$ (this expression is the definition of a uniaxial cluster),
$\psi_k$ is greater than $\psi_{\rm c}=\text{cos}\theta_{\rm c}$.
We note that
although we can theoretically consider multiaxial clusters with $\psi_k > \psi_{\rm c}$,
most multiaxial clusters meet the relation $\psi_k<\psi_{\rm c}$ in practice.
\newpage